\documentclass[twocolumn,iop]{emulateapj}

\slugcomment{Submitted to ApJ}

\shorttitle{Galactic disks do not need to survive in $\Lambda$-CDM}
\shortauthors{Puech et al.}

\begin{document}


\title{Galaxy disks do not need to survive in the $\Lambda$-CDM
  paradigm: the galaxy merger rate out to z$\sim$1.5 from
  morpho-kinematic data.}


\author{M. Puech\altaffilmark{1}, F. Hammer\altaffilmark{1}, P.F. Hopkins\altaffilmark{2}, E. Athanassoula\altaffilmark{3}, H. Flores\altaffilmark{1}, M. Rodrigues\altaffilmark{4,1,5}, J.L. Wang\altaffilmark{1}, and Y.B. Yang\altaffilmark{1,6}}
\affil{$^1$GEPI, Observatoire de Paris, CNRS-UMR8111, Univ. Paris-Diderot,
  5 place Janssen, 92195 Meudon, France}
\affil{$^2$Department of Astronomy, University of California, Berkeley, CA 94720, USA}
\affil{$^3$Laboratoire d'Astrophysique de Marseille, Observatoire
  Astronomique de Marseille Provence, Technopole de l'Étoile - Site de
  Chateau-Gombert, 38 rue Frédéric Joliot-Curie, 13388 Marseille Cedex
  13, France}
\affil{$^4$European Southern Observatory, Alonso de Cordova 3107 - Casilla 19001 - Vitacura -Santiago, Chile}
\affil{$^5$CENTRA, Instituto Superior Tecnico, Av. Rovisco Pais 1049-001 Lisboa , Portugal}
\affil{$^6$National Astronomical Observatories, Chinese Academy of Sciences, 20A Datun Road, Chaoyang District, Beijing 100012, China}




\begin{abstract}
  About two-thirds of present-day, large galaxies are spirals such as
  the Milky Way or Andromeda, but the way their thin rotating disks
  formed remains uncertain. Observations have revealed that half of
  their progenitors, six billion years ago, had peculiar morphologies
  and/or kinematics, which exclude them from the Hubble sequence.
  Major mergers, i.e., fusions between galaxies of similar mass, are
  found to be the likeliest driver for such strong peculiarities.
  However, thin disks are fragile and easily destroyed by such violent
  collisions, which creates a critical tension between the observed
  fraction of thin disks and their survival within the $\Lambda$-CDM
  paradigm. Here we show that the observed high occurrence of mergers
  amongst their progenitors is only apparent and is resolved when
  using morpho-kinematic observations which are sensitive to all the
  phases of the merging process. This provides an original way of
  narrowing down observational estimates of the galaxy merger rate and
  leads to a perfect match with predictions by state-of-the-art
  $\Lambda$-CDM semi-empirical models with no particular fine-tuning
  needed. These results imply that half of local thin disks do not
  survive but are actually rebuilt after a gas-rich major merger
  occurring in the past nine billion years, i.e., two-thirds of the
  lifetime of the Universe. This emphasizes the need to study how thin
  disks can form in halos with a more active merger history than
  previously considered, and to investigate what is the origin of the
  gas reservoir from which local disks would reform.
\end{abstract}





\section{Introduction}
Over the past years, observations and theory have been progressing
significantly in understanding how the first large disk galaxies could
form. Near-infrared (rest-frame optical) imaging from the HST reveals
that, while $\sim$40\% of z$\sim$1-3 most massive (i.e., $M_{stellar}
\geq 5\times10^{10}M_\odot$) galaxies are relatively compact and
quiescent, $\sim$60\% of them are actively forming stars and show
quite extended stellar structures (\citealt{weinzirl11}; see also
\citealt{vandokkum11,forster11}). In a larger range of mass (i.e.,
$\sim$10$^9$-10$^{11} M_\odot$), \cite{law11} suggested that the
stellar phase of z$\sim$1.5-3.6 galaxies is probably distributed as
triaxial ellipsoids rather than thick disks. Integral field
spectroscopy now routinely provides us with spatially-resolved
kinematics of the sub-population of z$\sim$1-3 emission line galaxies
\citep{forster06,bournaud08,vanstarkenburg08,wright09,forster09,law09,epinat09,lemoine10,lemoine10b,contini12,epinat12}.
Actively forming galaxies at these epochs reveal a relatively balanced
mix of obvious (major) mergers, relatively compact
dispersion-supported objects, and rotating extended structures, all
being much more turbulent in their gaseous phase compared to local
galaxies \citep{forster09}. On the theoretical side, cosmological
simulations have predicted that streams of cold gas would occur in the
halos in which these galaxies are expected to inhabit
\citep{dekel09,keres09,agertz09,brooks09,ceverino10,vandevoort11}.
Such streams would be responsible for bringing a continuous amount of
high angular momentum gas in these halos over a relatively short
timescale \citep{kimm11,pichon11}. Because the amount of mass brought
that way is expected to be considerable, this could in principle
pressurize the central forming disk and result in clump fragmentation
\citep{dekel09b,bournaud09,genzel11}, as ubiquitously observed at
these epochs \citep{elmegreen05,elmegreen09,elmegreen09b}. Such
``clumpy galaxies'' are now more and more considered as an important
step of galaxy formation. However, it remains unclear whether the
lifetime of such clumps is long enough to allow them to survive
feedback effects and coalesce to form a central bulge
\citep{elmegreen08,bournaud09b,murray10,krumholz10,genel10,forster11b,powell11,hopkins11}.
It remains also unclear how \emph{thin} disks can form from the high
angular momentum gas which is expected to be accreted into the halos.
Finally, the smoking gun detection of cold gas accretion is still
missing, although different options are being extensively discussed
\citep{dijkstra09,steidel10,goerdt10,benson11,faucher11,fumagalli11,kimm11b,letiran11,giavalisco11,stewart11,stewart11b}.

In contrast, it remains more uncertain how these z$\sim$2 thick
rotating disks evolved down to z$\sim$0, and how the local thin disks
formed. Indeed, the Hubble sequence of galaxies such as the Milky Way
($M_{stellar}=10^{10-11}M_\odot$) reveals a predominance (72\%) of
rotationally-supported disks in the present-day Universe
\citep{delgado10}. Using the Cosmological Principle, their progenitors
at z$\sim$0.6, i.e., six billion years ago, can be selected and their
morphology studied using the Advanced Camera for Survey on-board the
Hubble Space Telescope \citep{delgado10}. Using methodologies and
rest-frame luminosities that mimic those used for local galaxies, the
likeliest ancestor of the Hubble sequence can then be established
\citep{delgado10}. Half the progenitors of spiral galaxies show
peculiar morphologies that exclude them from this sequence
\citep{vandenbergh02,delgado10}. Most of them are starbursts, whose
emergence at z$\sim$0.6 reflects that of emission line galaxies
\citep{hammer97,mignoli05}. As part of the IMAGES survey, the
large-scale spatially-resolved kinematics of a representative
subsample of 63 emission line galaxies at z=[0.4-0.75] was obtained
using the GIRAFFE multi-IFU spectrograph
\citep{flores06,puech06,yang08}. Most peculiar galaxies also reveal
peculiar large-scale motions \citep{neichel08}. Comparing the
morpho-kinematic properties of the subsample of the 33 galaxies lying
into the CDFS \citep{yang08} with those of a grid of simple major
merger models, convincing matches were found in about two-thirds of
the cases \citep{hammer09b}. This leads to at least one third of the
progenitors of present-day spirals potentially involved in a major
merger. Since major mergers can easily destroy thin rotating disks
\citep{toomre72,toomre77,barnes88,barnes92,hernquist93,barnes96,toth92,hopkins08},
this creates a critical tension between the large fraction of
present-day disks and their survival within the $\Lambda$-CDM paradigm
\citep{stewart08,weinzirl09}.

Further investigating the ``disk survival issue'' requires an accurate
estimation of the evolution of the merger rate, i.e., the fraction of
galaxy involved in mergers per unit time. Several techniques were
developed over the past years to try to quantify accurately the
evolution of the merger rate, such as the pair technique (e.g.,
\citealt{lefevre00,kartaltepe07,lin08,rawat08,ryan08,bundy09,deravel09,deravel11})
or the morphological technique, either through non-parametric (e.g.,
\citealt{conselice03,lotz08,conselice09,shi09,lopez09}) or visual
(e.g., \citealt{jogee09,bridge10}) classifications. Alternatively, the
amplitude of the two-point correlation function can also be used to
infer a merger rate \citep{bell06,robaina10}. All these techniques are
very dependent on the visibility timescale of the specific merging
phase sampled by the observations, which requires to be estimated
using additional numerical simulations and/or analytical arguments
\citep{lotz08b}. Accounting for the exact distributions of merger mass
ratios or orbits involved is also challenging
\citep{lotz10a,lotz10,lotz11}. As a result, observational estimates
spread over almost one decade at z$\sim$0-1 \citep{hopkins09,lotz11}.

Recent progresses combine numerical simulations with transfer
radiation to derive a weighted average visibility timescale for each
observational technique, using predictions for the distribution of
mass ratios and other merger parameters from galaxy formation models
\citep{lotz11}. Taking these aspects into account, the discrepancy
between the different merger rates derived in the literature is
significantly reduced, while remaining differences are attributed to
differences in range of mass ratio and sample selection
\citep{lotz11}. This illustrates how crucial it is to carefully
accounts for these aforementioned effects. The present study puts this
effort one step forward: we show that by combining spatially-resolved
morphological and kinematic data as those provided by the IMAGES
survey, one can date back the epoch where observed galaxy mergers were
still in pair, and then derive the evolution of the fraction of
mergers out to z$\sim$1.5, in a consistent way as a function of
redshift. Such data also allow us to estimate relevant visibility
timescales that are by construction consistent with the distribution
of merger mass ratios and orbits of the galaxies in the observed
sample. The derived evolution of the merger rate out to z$\sim$1.5 is
found to be within a factor 2-3 of theoretical predictions from a
state-of-the-art $\Lambda$-CDM semi-empirical model.

This paper is organized as follows. Sect. 2 presents the data used in
this paper, which come from the IMAGES survey of z$\sim$0.6 emission
line intermediate-mass galaxies. In Sect. 3, we argue why large
morpho-kinematic perturbances most likely result from major mergers.
Sect. 4 presents the major merger fractions and rates derived from
these data. A comparison with state-of-the-art semi-empirical model of
galaxy formation is presented in Sect. 5. Finally, a discussion is
given in Sect. 6 and conclusions are drawn in Sect. 7. Throughout the
paper, magnitudes are given in the $AB$ system, and a WMAP-7 cosmology
is used \citep{komatsu11}, with ($\Omega _m$, $\Omega _\Lambda$,
$\Omega _b$, $h$, $\sigma _8$)=(0.275, 0.725, 0.0458, 0.702, 0.816).
We adopt the diet Salpeter IMF \citep{bell03} throughout this paper.
Note that all IMF-dependant quantities cited from other studies were
converted to this IMF.

\section{Morpho-kinematic data of z=[0.4, 0.75] galaxies: summary of the IMAGES survey}

\subsection{The representative  IMAGES-CDFS sample}
In the frame of the IMAGES survey, a sample of 35 emission line
galaxies at z=[0.4-0.75], i.e., about six billion years ago
(mean/median redshift of z=0.6/0.65), were observed in the Chandra
Deep Field South (CDFS) \citep{yang08}. This subsample of IMAGES is of
particular interest since it benefits from homogeneous imaging and 3D
data \citep{yang08,neichel08}. IMAGES targets were selected to have
$I_{AB}\leq$ 23.5 and $EW_0([OII])\geq 15\AA$ to guarantee their
detection by the multi-IFU spectrograph FLAMES/GIRAFFE at VLT
\citep{ravikumar07,flores06}. These galaxies were selected to be
intermediate-mass galaxies ($M_{stellar}=10^{10}-10^{11}M_\odot$ using
a diet Salpeter IMF and simplified prescriptions to convert J-band
luminosity and B-V colors into stellar mass, following
\citealt{bell03}). This range of mass is of particular interest since
most of the star formation at z$\leq$1 occurred in such galaxies
\citep{hammer05,bell05,zheng07}. Since the fraction of E/S0 did not
raise between z$\sim$0.6 and z=0 \citep{delgado10}, z$\leq$1
intermediate-mass galaxies are the likeliest progenitors of local
spirals (see also App. B).

From the full IMAGES sample, one galaxy was rejected because it turned
out not to be a starburst galaxy \citep{hammer09b,yang09}, and another
one because the HST/ACS images were corrupted and no morphological
analysis could be conducted. The resulting sample is fully
representative of the J-band luminosity function at z=0.5-1.0
\citep{yang08}. At these redshifts, the near-infrared luminosity
functions of the global population and of the blue sub-population of
galaxies are very similar \citep{cirasuolo07}. At first order,
emission-line and blue galaxies sample the same population
\citep{delgado10}, which in principle guarantees that the IMAGES-CDFS
sample is representative of the star forming population of galaxies at
these epochs.

\begin{figure}
\centering \includegraphics[width=9cm]{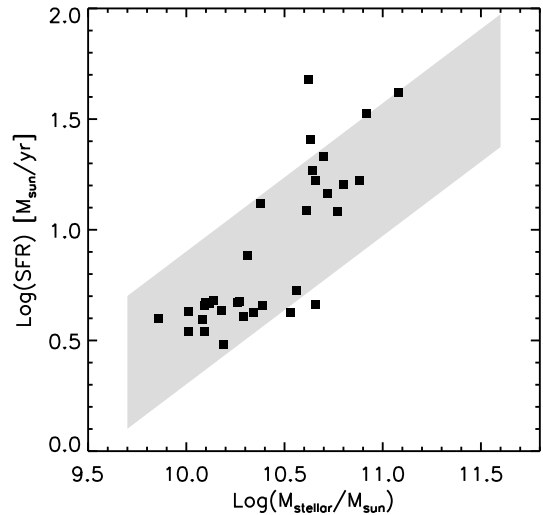}
\caption{The SFR-Stellar mass sequence in the IMAGES-CDFS sample
  (black squares). SFRs and stellar masses were rescaled to a diet
  Salpeter IMF. The grey region represents the z=0.45-0.7 SFR-Stellar
  mass sequence and its $\pm$1-$\sigma$ scatter from \cite{noeske07},
  as reported by \cite{dutton10}.}
\label{sfr}
\end{figure}

The total SFR of all the 33 remaining galaxies in the IMAGES-CDFS
sample was estimated as the sum of the unobscured SFR measured from
the rest-frame UV luminosity at 2800$\AA$ combined to the
dust-reprocessed contribution, which was estimated from the IR
luminosity as measured using 24 $\mu$m Spitzer fluxes (see
\citealt{puech10} for details). We find a mean SFR of
$\sim$12M$_\odot$/yr, which compares well with other UV+IR
measurements in larger samples at a similar range of mass and redshift
(e.g., \citealt{jogee09}). In Fig. \ref{sfr}, the location of the
IMAGES-CDFS galaxies in the SFR-M$_{stellar}$ plan is compared to the
sequence of starbursting galaxies observed by \cite{noeske07} at
z=0.45-0.7. As expected, the IMAGES-CDFS sample comprises Luminous
InfraRed Galaxies, which populates the high-SFR tail of the
distribution at SFR$\geq$14M$_\odot$/yr \citep{lefloch05}, and less
obscured (UV) starbursts, which populate the low-SFR tail of the
distribution. This figure reveals that most (91\%) IMAGES-CDFS
galaxies lie well within the observed SFR-M$_{stellar}$ sequence at
these redshifts. The star formation rate density in the IMAGES-CDFS
sample is found to be $\log{(\rho _{sfr})}\sim$-1.4 (where $\rho
_{sfr}$ is expressed in $M_\odot.yr^{-1}.Mpc^{-3}$), also in good
agreement with expectations from larger surveys that include IR data
at 24 $\mu$m \citep{zheng07}.\footnote{Apparent discrepancies of up to
  a factor two can appear in comparison with other studies based on
  different survey areas, stellar mass cuts, and methods for
  estimating the SFR. For instance, \cite{jogee09} found $\log{(\rho
    _{sfr})}=$-1.4 at z$\sim$0.6 for galaxies with
  M$_{stellar}\geq10^{9.1}$M$_\odot$ and $\log{(\rho _{sfr})}=$-1.9
  for galaxies with M$_{stellar}\geq10^{10.5}$M$_\odot$ (using a diet
  Salpeter IMF). Using a larger area survey, \cite{zheng07} found that
  for the mass range M$_{stellar}=10^{10.15}-10^{11.15}$M$_\odot$,
  i.e., very close to the range of mass considered in the present
  study, $\log{(\rho _{sfr})}$ ranges from -1.56$^{+0.07}_{-0.08}$ at
  z=0.4-0.6 to -1.34$\pm$0.05 at z=0.5-0.8, which is consistent with
  the value found in the IMAGES-CDFS sample.} This confirms that the
IMAGES-CDFS sample is representative of star-forming galaxies at
z$\sim$0.6.

Gas masses were estimated from the total SFR by inverting the
Schmidt-Kennicutt law \citep{puech10}. All galaxies were found in a
range of total baryonic mass $\log{(M_b/M_\odot)}=$10.2 to 11, or
$\log{(M_{stellar}/M_\odot)}$ in the range 10 to 11. The stellar-mass
density in the IMAGES-CDFS sample is found to be $\log{(\rho
  _{stellar})}\sim$8.1 (where $\rho _{stellar}$ is expressed in
$M_\odot.Mpc^{-3}$), which is in relatively good agreement with
expectations from larger surveys of blue galaxies
\citep{arnouts07,bell07}, within expected uncertainties.

\subsection{Morpho-kinematic classification}
Detailed color morphology (at a $\sim$500 pc resolution scale) and
large-scale kinematics (at a $\sim$7 kpc resolution scale) were
obtained thanks to the Hubble Space Telescope, and 3D spectroscopy
using the GIRAFFE multi-IFU spectrograph, respectively. Galaxies were
classified independently according to their morphology
\citep{neichel08} and kinematics \citep{yang08}, based on similarities
with the properties of local galaxies.

The kinematical maps allowed us to classify the IMAGES-CDFS galaxies
into three distinct kinematic classes (see \citealt{yang08}). Rotating
disks [RDs] were those for which the Velocity Field [VF] showed an
order gradient, a dynamical axis aligned with the morphological axis,
and a velocity dispersion peak located at the dynamical center. For
these objects, it was required that the sigma peak location and
amplitude can be reproduced by modeling the velocity dispersion map
from the velocity gradient observed in the VF. At the GIRAFFE spatial
resolution, it is indeed expected that most of the rotation curve
gradient falls into on GIRAFFE pixel, which results in an velocity
dispersion peak at the dynamical center of the object. Conversely,
galaxies with complex kinematics [CKs] were those for which no ordered
gradient was observed in the VF, or when a significant misalignment
between the kinematical rotation axis and the morphological principle
axis was detected. Perturbed rotators [PRs] gathered objects with an
order velocity gradient in the VF, but for which the velocity
dispersion peak (location and/or amplitude) could not be reproduced
from the VF, which indicates that this peak cannot be attributed to
rotation. This classification was tested against quantified criteria
and considering all uncertainties associated to kinematic
measurements, which makes it objective and reproducible, as described
extensively in \cite{yang08}.

The detailed color morphology of all the 33 galaxies in the
IMAGES-CDFS sample was studied by \cite{neichel08}. Using a
reproducible decision tree relying on parametric parameters and visual
inspection of images and color maps, galaxy morphologies were
classified into six different classes: Compact systems, Mergers,
Peculiar/ Possible Mergers, Peculiar Tadpoles, Peculiar Irregulars,
and Spiral disks. In particular, it was required for a rotating disks
to show redder colors towards the center as observed in local
galaxies. This classification was done by three independent
classifiers using a decision tree, which limits the subjectivity and
makes it reproducible. The comparison between the morphological and
kinematical classes show a very good agreement, with more than 80\% of
spiral disks having rotating disks, and more than 90\% of peculiar
galaxies having complex or perturbed kinematics.

Since perturbations in the (ionized) gaseous phase generally
correspond to perturbations in the stellar phase, the result of both
the morphological and kinematical classifications can be robustly
synthesized into a morpho-kinematic classification \citep{hammer09b}.
With this classification, rotating spiral disks are those classified
both as rotating disks from the kinematic point of view, and as spiral
disks from the morphological point of view. Non-relaxed systems are
those that show a complex kinematics and a peculiar morphology.
Semi-relaxed systems are galaxies that show either some rotation from
their kinematics (being classified as either a rotating disk or either
a perturbed rotation) but with a peculiar morphology, or galaxies that
do not show rotation in their kinematics (being classified as having a
complex kinematics) but with a spiral disk morphology. According to
this classification, 80\% of z$\sim$0.6 starburst galaxies are
non-relaxed or semi-relaxed systems \citep{hammer09b}, which
corresponds to $\sim$50\% of all intermediate-mass, z$\sim$0.6
galaxies, since 60\% of intermediate-mass, z$\sim$0.6 galaxies are
indeed starbursts \citep{hammer97,mignoli05}. The HST/ACS images and
kinematic maps for a representative sub-sample of the IMAGES-CDFS can
be found in Fig. \ref{vf} (see App. A).

\section{Associating kinematic disturbances with physical processes}
Stellar kinematic disturbances are indubitable evidence for an
out-of-equilibrium dynamical state. In high-redshift galaxies,
deriving spatially-resolved stellar kinematics remains out of reach,
and one has to rely on gas kinematics as a best proxy. This implicitly
assumes that gas kinematics traces the gravitational potential, and
that no other mechanisms are perturbing gaseous motions, such as,
e.g., outflows. In the IMAGES sample, large-scale outflows have been
detected in only a handful of cases \citep{puech10}. Using FORS2 slit
spectroscopy, \cite{rodrigues11} further estimated that only 8\% of
intermediate-mass galaxies at z$\sim$0.6 have significant outflows.
This suggests that gas kinematics is indeed a reliable tracer of the
dynamical state of galaxies, at least on large spatial scales. Several
dynamical processes are known to be a source of kinematic
disturbances. Some of them are local in space, while others are
global. Relative to the equilibrium state of rotation (as expected in
the intermediate mass range since more massive galaxies are expected
to reach a different equilibrium state supported by dispersion), local
disturbances are either due to minor mergers or internal instabilities
such as clump formation, while global disturbances can be due to major
mergers or cosmological gas accretion \citep{hopkins10}.

\subsection{Local disturbances}
\begin{figure*}
\centering \includegraphics[width=18cm]{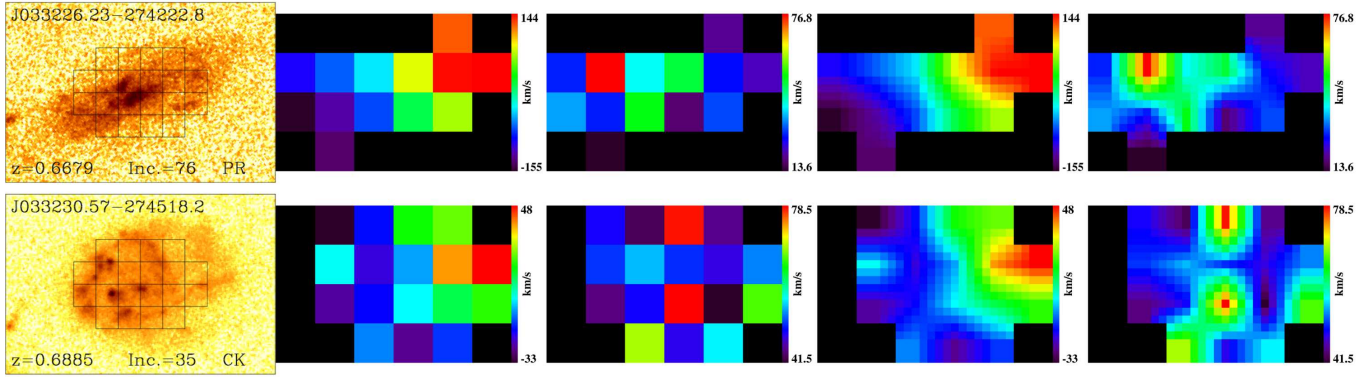}
\caption{Examples of large-scale kinematic maps obtained in the frame
  of the IMAGES-CDFS survey with FLAMES/GIRAFFE at VLT.\emph{From left
    to right:} HST/ACS z-band image superimposed with the GIRAFFE IFU
  grid (0.52 arcsec/spaxel), velocity field, velocity dispersion map,
  5$\times$5 linearly interpolated velocity field, and 5$\times$5
  linearly interpolated velocity dispersion map. \emph{Top panel:}
  J033226.23-274222.8, which is an example of a minor merger detected
  at z$\sim$0.67 (see \citealt{puech07}). \emph{Bottom panel:}
  J033230.57-274518.2, which is an example of a z$\sim$0.69 clumpy
  galaxy with Complex Kinematics. The local perturbations due to minor
  mergers or local instability remain undetected in the velocity field
  at the GIRAFFE spatial resolution (see text).\label{figgals}}
\end{figure*}

In the IMAGES sample, kinematic perturbations are detected over a
$\ge$7 kpc spatial scale, which corresponds to the GIRAFFE-IFU spatial
resolution element (i.e., two spaxels). They generally affect the
galaxies across a significant fraction of their diameter, which, in
principle, excludes \emph{de facto} any mechanism resulting in local
kinematic disturbances. In order to quantify this, we derived the
Jeans scale length $L_J$ \citep{bournaud09} in the IMAGES-CDFS, which
is found to be 0.7$\pm$0.1 kpc. $L_J$ increases to 1.0$\pm$0.4 kpc in
the sub-sample of clumpy galaxies that show rotation \citep{puech10b}.
Thus, such Jeans instabilities should remain largely undetected at the
GIRAFFE IFU spatial resolution. In the bottom of Fig. \ref{figgals}
one of the most clumpiest galaxy in the IMAGES-CDFS sample. This
galaxy shows at least seven clumps spread over its disks and is
classified as CK. However, the kinematic perturbations are mostly
large-scale, with a misalignment between the morphological and
kinematic axis as defined by the location of the maximal and minimal
rotation velocities, and with a velocity dispersion peak that cannot
be reproduced by the rotation model inferred from the VF
\cite{yang08,puech10}. This confirms that even in the most disturbed
clumpy galaxies, the kinematic perturbations cannot be associated with
the position of the clumps, i.e., they are not local but global. In
addition, clump instabilities are limited to a fraction of 20\% of
intermediate-mass galaxies at z$\sim$0.6 that are effectively clumpy
\citep{puech10b}, which also rules out clump instabilities as a driver
for the strongest kinematic perturbations.

Regarding minor mergers, such processes are also expected to lead to
local kinematic perturbations only. An obvious case of minor merger
with a mass ratio of $\sim$1:18 was studied in detail by
\cite{puech07}: only a local signature is found in the velocity
dispersion map (see top panel in Fig. \ref{figgals}). Even if we
cannot exclude that in very specific cases such local processes could
result in large kinematic disturbances such as those observed in
IMAGES galaxies, we can safely conclude that they cannot be the main
driver for the majority of galaxies, at least for the CK galaxies.
This does not mean that minor mergers are not ongoing in the IMAGES
sample, but given the spatial resolution of GIRAFFE-IFU data, they
probably remain mostly undetected.

\subsection{Global disturbances}
Regarding the impact of processes resulting in global disturbances, it
is instructive to consider the range of the relative variations of the
gravitational potential (compared to the dynamical equilibrium state)
$\Delta \phi/\phi$, in which galaxies can be considered as isolated
and subject to secular evolution, as a function of mass and redshift
\citep{hopkins10}. We estimated $\Delta \phi/\phi$ for galaxies in the
IMAGES-CDFS sample using rotation kinematic models \citep{puech08}.
For this, we derived the pixel-by-pixel relative difference between
the observed VF and each galaxy model, and considered the sum $\Delta
V/V$ of this map over all pixels, weighted by the observed S/N in each
pixel. The quantity 2.$\Delta V/V$ can then be used as a proxy for
$\Delta \phi/\phi$. We also derived the dynamical timescale $t_{dyn}$
for each galaxy in the IMAGES-CDFS sample using as a proxy the ratio
between the gas radius and the rotation velocity as determined by
\cite{puech10}. In galaxies showing rotation (RDs+PRs), we found that
$\Delta \phi/\phi$=1.2$\pm$0.4 (median value) and $t_{dyn}$=60$\pm$3
Myr, while for CK galaxies, $\Delta \phi/\phi$=3.1$\pm$1.0 and
$t_{dyn}$=60$\pm$24 Myr. Uncertainties correspond to 1-$\sigma$
error-bar, and were estimated using bootstrap resampling.

In this regime (i.e., $t_{dyn} \ga$10 Myr and $\Delta \phi/\phi
\ga$1), the two dominant physical processes expected to drive
gravitational perturbations in this range of mass and redshifts are
secular evolution (i.e., instabilities occurring in a galaxy) and the
direct dynamical evolution of individual substructures due to some
previously initiated perturbation, which can eventually dominate the
evolution of the whole system \citep{hopkins10}. Large global
perturbations (i.e., $\Delta \phi/\phi \ga$1) are typically triggered
by major mergers. The characteristic timescale for secular evolution
can be estimated using the circular disk period i.e., $P_d=2\pi
R/V_c$, with $R$ the disk radius and $V_c$ the circular velocity. We
used again gas radius and the rotation velocity as determined by
\cite{puech10} as proxies for these two quantities. The timescale for
dynamical friction associated with the dynamical evolution of
substructure was estimated following the simplified prescription given
by \cite{hopkins10}:
$$
t_{df}=\frac{0.2}{\delta _0 ln(\Lambda)}P_d,
$$
where $\delta _0=\Delta \phi/\phi$, and $\Lambda$=1+1/$\delta _0$.
This timescale quantifies how fast the dynamical evolution of
substructures evolve on their own, once they have been triggered. For
instance, a merging galaxy will trigger instabilities in the main
progenitors that will subsequently evolve on their own, on a timescale
given by $t_{df}$ \citep{hopkins10}. The dominant process at given
amplitude of perturbation is the one associated with the shortest
timescale.

\begin{figure}
\centering \includegraphics[width=9cm]{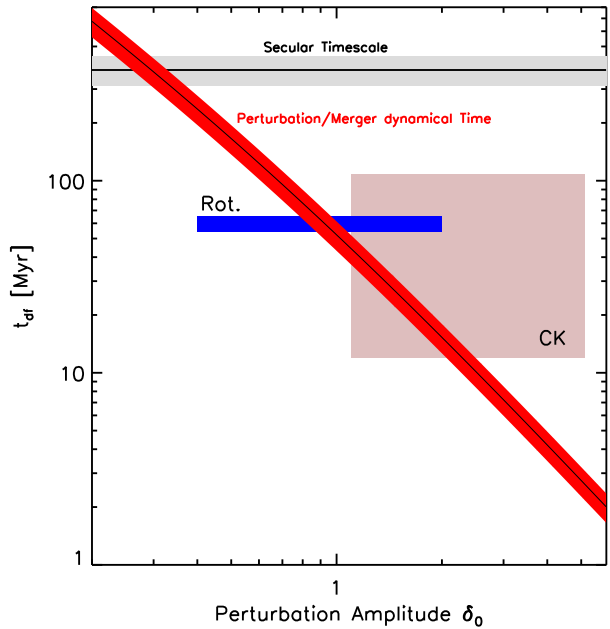}
\caption{Dynamical timescales (in Myr) for the evolution of secular or
  self-amplified perturbations as a function of the amplitude of the
  gravitational perturbation. The grey region represents the expected
  2-$\sigma$ uncertainty range around the median timescale for secular
  perturbations, while the red region represents the expected
  2-$\sigma$ uncertainty range around the median timescale for
  individual perturbations to evolve on its own (as approximated by
  Eq. (5) of \citealt{hopkins10}, see text). The blue region is the
  2-$\sigma$ uncertainty range around the median value obtained for
  rotating objects at z$\sim$0.6, while the brown region represents
  the same for objects with a complex kinematics.\label{figpot}}
\end{figure}

In Fig. \ref{figpot}, we plotted the expected range of value of these
two regimes at z$\sim$0.6 in the IMAGES-CDFS sample. Secular evolution
is found to be dominant at small relative variations of the
gravitational potential $\Delta \phi/\phi$, while the influence of
(major) mergers dominate above $\Delta \phi/\phi \geq$0.3. More
specifically, isolated galaxies should be found in the upper left
region of Fig. \ref{figpot}, close to the grey region (as well as
large local disks such as M31), while on-going mergers and mergers
remnants should be found at $\Delta \phi/\phi \sim$1 and close to the
red region. Comparing with these theoretical expectations, CK galaxies
lie closer to the individual dynamical perturbation regime (i.e., the
merger-driven regime), which is consistent with (major) merger-driven
kinematic disturbances in these galaxies. Galaxies showing rotation
also lie in this regime, although with a smaller range of
gravitational perturbation, which is consistent with these galaxies
being more (temporally) advanced merger remnants, as suggested by
their other morpho-kinematic properties (see \citealt{hammer09b} and
Sect. 4.1). On overall, z$\sim$0.6 rotators (RDs+PRs) lie closer to
the expected merger-driven regime, which probably reflects the fact
that they are closer to dynamical equilibrium than CK galaxies, in
which gravitational perturbations are about three times larger. This
suggests that any other known process (gas accretion or secular
evolution) is very unlikely to be responsible for the high level of
kinematic disturbance observed on large spatial scales, at least for
the majority of the IMAGES-CDFS sample.

\subsection{Comparison with major merger models}
In this context, it was explored whether hydrodynamical major merger
models for the 27 galaxies in the IMAGES-CDFS sample which are not
relaxed morphogically and/or kinematically could generally account for
their perturbations
\citep{hammer09b,hammer09,puech09,peirani09,fuentes10}. Morphological
and kinematic maps from these simulations were produced and degraded
to the spatial resolution of the observations. Such maps were then
explored visually as a function of time and viewing angles to
determine the best match to observations. The so-determined best model
was graded by three different examiners (independent of the one who
determined the best model) using specific constraints
\citep{hammer09b}. Secure major merger candidates (18/27) were
identified as galaxies having a model that fitted observations with a
good level of confidence (i.e., grades $\geq$ 4/6). Examples of such
models are shown in App. A. These candidates can be used as a basis to
estimate the fraction of mergers and the merger rate at different
epochs as detailed in Sect. 4.1. Interestingly, the modeling exercise
lead to a median merger mass ratio $\mu$=3.03$\pm$0.34 in the
IMAGES-CDFS sample, while theoretical expectations from variations of
the gravitational potential give consistent predictions with $\Delta
\phi/\phi \sim \mu$=3.1$\pm$1.0 in CK galaxies (see above). We refer
the reader to \cite{hammer09b} for more details about the comparison
between simulations and observations. This comparison suggests that at
least 33\% of z$\sim$0.6 galaxies (50\% of galaxies not relaxed
morpho-kinematically according to the morpho-kinematic classifications
$\times$ the ratio of secured models, which is 18/27=66\%) are
involved in major mergers with a median mass ratio $\sim$3.

\section{Deriving the merger rate out to z$\sim$1.5 from morpho-kinematic observations}

\subsection{Methodology}
\begin{figure*}
\centering \includegraphics[scale=0.7]{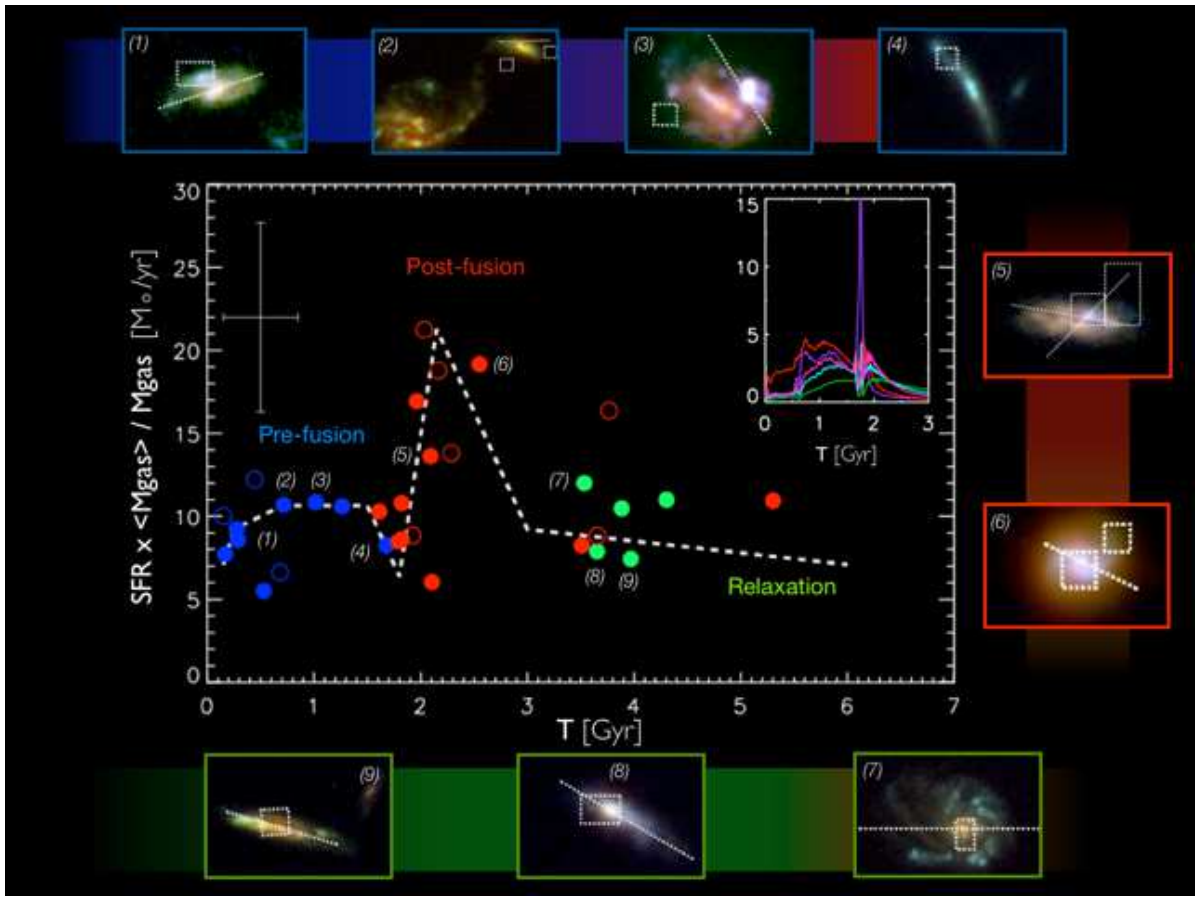}
\caption{Time sensitivity of morpho-kinematic observations to the
  major merger process. Galaxies are plotted at the time at which
  individual numerical models give the best-fit to observations. Full
  symbols represent galaxies whose morpho-kinematics compared best
  with those of their models. Merging galaxies were classified into
  three different classes: the pre-fusion (blue symbols), post-fusion
  (red), and relaxation phases (green). The pre-fusion phase
  corresponds to major mergers where both progenitors can be visually
  identified, while the post-fusion phase corresponds to those where
  the fusion is too advanced to reliably distinguish them. The dash
  line is not a fit but a simple visual guide through the points. The
  median uncertainty is indicated is the upper-left corner. The
  upper-right inset shows typical similar curves as reproduced from
  hydrodynamical simulations of major merger (\citealt{cox06}).
  Surrounding stamps show IMAGES-CDFS HST/ACS B-V-z color images with
  the approximate location of the kinematic axes (dash white lines)
  and velocity dispersion peaks (white dash boxes) superimposed. These
  stamps are numbered in increasing time sequence: (1)
  J033224.60-274428.1, (2) J033220.48-275143.9, (3)
  J033239.72-275154.7, (4) J033227.07-274404.7, (5)
  J033230.43-275304.0, (6) J033245.11-274724.0, (7)
  J033237.54-274838.9, (8) J033241.88-274853.9, and (9)
  J033212.39-274353.6. Their morpho-kinematic observations are shown
  in Fig. \ref{vf} as well as the result of their modelling by
  \cite{hammer09b}, see Appendix A.\label{figtime}}
\end{figure*}

All the merger phases that are expected to lead to a significant SFR
enhancement, i.e., the pre-fusion, fusion/post-fusion, and relaxation
phases \citep{larson78}, are homogeneously populated as a function of
time by the IMAGES-CDFS galaxies, as evidenced by Fig. \ref{figtime}.
In this figure, SFRs were homogenized by dividing the observed SFR by
the estimated mass of gas in each galaxy and multiplying each derived
number by the average mass of gas in the sample. The time sequence
reproduces quite well expectations from modeling the star formation
during a typical merger (see insert). This shows that morpho-kinematic
data such as those provided by the IMAGES survey are sensible to all
the phases of a gas-rich major merger event.

Amongst the galaxies populating the relaxation phase, six of them are
rotating, but they are clearly distinct of local rotating disks
because of their high average star formation rate, and higher gas
velocity dispersion \citep{puech07,epinat10}. The latter suggests that
these galaxies are possibly observed during a post major merger
relaxation phase (\citealt{robertson08,monreal10}; see also Fig.
\ref{figpot} and Sect. 3.2). Most of the star formation activity is
indeed concentrated at the disk outskirts, which suggests an ongoing
inside-out growth of their disks \citep{neichel08,hammer09b}. The time
sequence in Fig. \ref{figtime} empirically accounts for a cosmological
distribution of merger parameters, since it is constructed using a
representative sample. It can thus be used to estimate accurately the
evolution of the merger fraction and merger rate as follows.

Theory usually gives the merger rate at the time when halos start
merging, i.e., where galaxies that inhabit these halos are still in
pair. The epoch when a given merging system was still a pair is dated
using individual numerical models from \citet{hammer09b}. An
homogeneous and self-consistent determination of the merger rate as a
function of lookback time can be estimated by simply counting the
fraction of galaxies in each phase and dividing this number by the
corresponding average visibility timescale provided by the modeling of
IMAGES galaxies. This method assumes that the visibility timescales do
not evolve with redshift. Such an assumption is consistent with
semi-analytic models, which predict almost no evolution of the merger
visibility timescale as a function of lookback time, at least within
uncertainties \citep{kitzbichler08}. Another possible caveat is the
observed evolution of the galaxy gas fraction with redshift
\citep{puech10,daddi10,rodrigues11,erb06}, which could potentially
result in evolving timescales depending of the merging phase probed
\citep{lotz10}. However, hydrodynamical simulations of binary mergers
\citep{lotz10} predict that unequal mass ratio mergers (which dominate
the empirical distribution of merger mass ratios, see Sect. 3.2)
generally show no evolution of the pair visibility timescale with gas
fraction.

\subsection{The fraction of mergers out to z$\sim$1.5}

\subsubsection{Pre-fusion and post-fusion phases} 
Depending on whether or not the two progenitor candidates potentially
involved in the merger could be identified visually in the HST/ACS
images (e.g., when two candidates for the progenitor nuclei could be
identified), the secured major merger candidates were split into
pre-fusion (8 gal.) or post-fusion phases (10 gal.). Amongst the
objects that could not be secured as major merger candidates using
numerical models, three were found to correspond to the pre-fusion
phase, since they reveal distinct potential progenitors. They were
therefore added to the pre-fusion phase. We attribute the fact that
their numerical models did not succeed in reproducing their
morpho-kinematical properties with sufficient precision to the limited
exploration of the parameter space.

The pre- and post-fusion phases as observed at z$\sim$0.6 correspond
to galaxies that were on average in pairs (and before the first
passage) at earlier epochs. To date these epochs, we simply averaged
the time elapsed since the beginning of the merger model for each of
the 18 secured major merger candidates, and converted these numbers
into a redshift. All numbers were weighted by the respective grade
given to the different models. It was found that the pre-fusion phase
corresponds on average to z$\sim$0.72$^{+0.10}_{-0.04}$ (i.e.,
0.7$\pm$0.2 Gyr earlier than z=0.6), while the post-fusion phase
corresponds on average to z$\sim$1.12$^{+0.13}_{-0.10}$ (i.e.,
2.5$\pm$0.4 Gyr earlier than z=0.6). Uncertainties were estimated
using bootstrap resampling.

The fraction of mergers in emission-line galaxies at z=0.4-0.75 for
the pre-fusion phase is 0.5x11/33 = 16.7\% (with the factor 0.5
accounting for the fact that only one galaxy in the pair is part of
the sample), while the fraction of mergers in emission-line galaxies
for the post-fusion phase is directly given by 10/33 = 30.3\%. Since
60\% of intermediate-mass galaxies at z$\sim$0.6 are emission-line
galaxies \citep{hammer97,mignoli05}, this translates into a total
fraction of mergers of 10.0\% and 18.1\% for the pre- and post-fusion
phases, respectively. This assumes that major mergers produce a star
formation activity that results in $EW_0([OII])\geq 15\AA$. However,
it has been shown that major mergers can lead to modest SFR
enhancement over relatively short timescales
\citep{dimatteo08,cox08,robaina09,jogee09}. This assumption can
therefore lead to possible systematic effects that are considered
separately (see Sect. 4.4). Random uncertainties due to statistical
fluctuations in the respective sub-samples (see Tab. \ref{rates}) are
derived accurately using a beta distribution generator as an estimator
of confidence intervals for the binomial statistics \citep{cameron10}.

The 18 secured pre- and post-fusion major merger models were used to
determine the fusion time for each galaxy. The fusion time was defined
as the interval of time between the beginning of the simulation (i.e.,
when the two progenitors are in pair and before the first passage) and
the time at which the two nuclei merge according to the best-fit
model. These individual timescales were averaged to mitigate
uncertainties due to statistical fluctuations in the respective
sub-samples. The averaged pre-fusion phase is found to be 1.8$\pm$0.1
Gyr long, which gives an estimate of the pre-fusion visibility
timescale in the IMAGES-CDFS sample. The uncertainty was derived by
bootstrap resampling of the grade-weighted distribution of individual
timescales. We assume the same visibility timescale for the
post-fusion phase, since both appear to be similar in the IMAGES
sample, according to Fig. \ref{figtime}.

We compare this estimate with hydrodynamical simulations including
radiation transfer \citep{lotz08}. We selected the G2 and G3 models as
being the closest to the IMAGES-CDFS galaxies in terms of baryonic
mass and gas fractions. Indeed, the most important parameters driving
the result of a major merger are the baryonic mass (as a proxy for the
total mass inside the optical radius) and the gas fraction
\citep{hopkins09c}. For these two models, the timescale from the
beginning of the interaction to the fusion of the two nuclei is found to
be 1.24 and 2.44 Gyr respectively, with an average $\sim$1.84 Gyr, in
good agreement with our estimate\footnote{The post-fusion duration
  was not measured in these simulations but set to an arbitrary time
  after the coalescence of the nuclei, which prevents any further
  comparison.} of 1.8$\pm$0.1 Gyr. We also compared the pre-fusion
visibility timescale with an average pair visibility timescale for
large projection distances (i.e., $R_{proj} \leq 100$ kpc$.h^{-1}$)
derived from similar simulations \citep{lotz10a,lotz10}, which is
found to be 1.81$\pm$0.06 Gyr. Taking into account predictions of mass
ratios from galaxy formation models slightly reduces the average value
to 1.5-1.6 Gyr \citep{lotz11}. The pre-fusion phase includes galaxies
that are too close to be selected as pairs by usual criteria (see also
Sect. 4.3). It is therefore expected that the pre-fusion visibility
timescale in the IMAGES-CDFS sample is slightly larger than the
average pair visibility timescale.

\subsubsection{Relaxation phase}
The six rotating spiral disks in the IMAGES-CDFS can be used to
estimate the fraction of mergers at even earlier epochs (see Sect. 3.2
and 4.1). Their mass doubling timescale $T_{sfr}$ is provided by the
ratio between their total SFR and stellar mass \citep{neichel08}. Note
that this timescale is independent of the IMF. One of them shows a
very large $T_{sfr}$ ($\sim$7 Gyr) and was therefore considered as an
interloper and discarded in the following. Assuming that these objects
are merger remnants, one can attribute most of their star formation
activity to the formation of the disk (see Sect. 4.1 and 6.2).
Therefore, this timescale can be used to roughly estimate the average
time elapsed from fusion to the observation epoch of these galaxies.
On average, these galaxies were therefore observed as pairs at a
lookback time 1.8+1.8 (pre-fusion + post-fusion phases) +
0.5x($<T_{sfr}>$-1.8) = 3.9$\pm$0.3 Gyr, with $<T_{sfr}>$=2.3$\pm$0.3
Gyr (where the error-bar was estimated using bootstrap resampling).
This corresponds to z$\sim$1.55$^{+0.15}_{-0.13}$. The average
relaxation timescale is found to be 5.5-1.8-1.8=1.9 Gyr (see Fig.
\ref{figtime}).

Amongst the nine non- or semi-relaxed galaxies in the IMAGES-CDFS
sample that could not be securely identified as major mergers, six
were found to be in the post-merger phase according to their best
models. They were all modeled quite at the end of the 3 Gyr duration
of the simulations, which suggests that the morpho-kinematics of these
galaxies might not have been identified as resulting from a major
merger due to the limited duration of the simulations. Amongst them,
four show evidence for rotation in their velocity fields and were
classified as perturbed rotators, which is consistent with this
interpretation. We therefore included them as part of the relaxation
phase (see red points in the relaxation phase in Fig. \ref{figtime}).
This leads to a fraction of mergers at z$\sim$1.6 of 20.0\% (see Tab.
\ref{rates}).

\subsection{Merger rates}
The merger rate r [Gyr$^{-1}$] corresponding to the three merger
phases considered above (i.e., the pre-fusion, post-fusion, and
relaxation phases) can be estimated using the ratio between the
corresponding merger fraction and visibility timescale (see Tab.
\ref{rates}). Error-bars due to random fluctuations in the sub-samples
were estimated by propagating the random uncertainties on the fraction
of mergers and on the visibility timescale. The derived merger rates
and their associated uncertainties are shown in Fig. \ref{figrate}.

\begin{table*}
\begin{center}
  \caption{Galaxy merger rate derived from the IMAGES-CDFS survey.
    \emph{From left to right:} Merger fraction, merger rate, and
    redshifts sampled by the IMAGES-CDFS survey. Quoted uncertainties
    correspond to 1-$\sigma$ random and systematic error-bars
    respectively.\label{rates}}
\begin{tabular}{cccc}
\tableline\tableline
Merger phase & Fraction of mergers (\%) & Merger rate (\%.$Gyr^{-1}$) & Redshift\\\tableline
Pre-fusion   & $10.0^{+5.8}_{-2.6}$$^{+3.3}_{-3.7}$ & $5.5^{+3.5}_{-1.8}$$^{+1.9}_{-2.0}$  & $z=0.72^{+0.10}_{-0.04}$\\
Post-fusion  & $18.1^{+6.7}_{-3.8}$$^{+3.3}_{-5.5}$ & $10.1^{+4.3}_{-2.7}$$^{+1.8}_{-3.3}$ & $z=1.12^{+0.13}_{-0.10}$\\
Relaxation   & $20.0^{+6.8}_{-4.0}$$^{+3.3}_{-9.1}$ & $11.1^{+4.4}_{-2.7}$$^{+1.8}_{-5.1}$ & $z=1.55^{+0.15}_{-0.13}$\\
\tableline\tableline
\end{tabular}
\end{center}
\end{table*}

For selection criteria similar to those of the IMAGES-CDFS sample, a
fraction of galaxies in pair $\sim$2-6\% at z$\sim$0.6 is generally
found \citep{rawat08,kartaltepe07,bundy09}. In the present study, the
fraction of galaxies involved in major mergers is found to be
10$\pm$3\% at z$\sim$0.7. Such a fraction is larger than the fraction
of galaxies in pairs since the pair technique only picks up
progenitors when they are approaching for the first time and do not
consider progenitors between the first and second passage and/or when
they are getting too close one from each other, while morpho-kinematic
data are sensitive to all phases. This is reflected in the visibility
timescale associated with pair fractions, which is often assumed to be
$\sim$0.5 Gyr from analytic arguments or independent simulations
(i.e., which do not necessarily reflect the exact properties of the
observations; see also \citealt{lotz11}). Using such a timescale and a
fraction of galaxies in pair $\sim$2-6\% indeed leads to a merger rate
$\sim$4-12\% Gyr$^{-1}$, which is consistent with our z$\sim$0.7
estimate of 5.5\%. When no lower limit to the distance between the two
progenitors is imposed, the pair technique gives very similar results,
with $\sim$9\% of galaxies in pair at z$\sim$0.75, and $\sim$15\% at
z$\sim$1 \citep{ryan08}.

The morphological technique provides estimates of the fraction of
morphologically-disturbed mergers of 7-11\% at z$\sim$0.7, 11-17\% at
z$\sim$1.1, and 27$\pm$11\% at z$\sim$1.4 for similar mass ranges and
depending and the method used (CAS/GM system, see
\citealt{conselice08,lotz08,conselice09}, or visual inspection, see
\citealt{bridge10,jogee09}). Within uncertainties, this is consistent
with the fraction of mergers found using morpho-kinematic data (see
Tab. \ref{rates}).

We conclude that alternative techniques generally give consistent
estimates, although with a resulting large range of values
\citep{hopkins09}, which reflects systematic effects due to their
limited sensitivity to the merging process and different sample
selection effects (see also \citealt{lotz11}).

\subsection{Systematic uncertainties}
A first possible cause of systematic effects is linked to the
selection criteria of the IMAGES-CDFS sample, and in particular to the
criterion $EW_0([OII])\geq 15\AA$. Indeed, we assumed that all
quiescent ($EW_0([OII])< 15\AA$) galaxies are not major mergers,
whereas a significant fraction of them (25\%) were found to have
peculiar morphologies \citep{delgado10}, which could be linked to less
active mergers. Indeed, numerical simulations have revealed that major
mergers do not always lead to strong enhancements of the star
formation rate \citep{dimatteo08}, and therefore these peculiar
quiescent galaxies might also be undergoing major merger events. This
could potentially rise the fraction of mergers by 10.0\% since
quiescent galaxies represent 40\% of intermediate-mass galaxies at
z$\sim$0.6 \citep{delgado10}. Since we do not have kinematic data for
these objects, this could potentially affect any of the three merger
phase considered in this study. We therefore considered that, on
average, the merger fraction in each phase could be affected by a
$\sim$3.3\% systematic effect upward (see Tab. \ref{rates}).

The number of mergers in the pre-fusion phase could also be
overestimated due to fly-by interlopers. We estimated that four out of
the eleven objects in this phase could be possible fly-byes and escape
a subsequent merger (J033217.62-274257.4, J033219.32-274514.0,
J033220.48-275143.9, J033238.60-274631.4). These objects correspond to
cases where the two potential progenitors are well separated and
photometric redshifts are consistent, but no reliable spectroscopic
information was available to confirm whether or not the two galaxies
were indeed gravitationally bound. This can potentially have an
impact on the fraction of mergers in the pre-fusion phase by 3.7\%
downward, and the corresponding merger rate by 2.0\% (see Tab.
\ref{rates}).

Six galaxies in the IMAGES-CDFS are consistent with expectations from
clumpy galaxies \citep{puech10b}, i.e., gas-rich Toomre-unstable
disks. These six galaxies are found to be in the post-fusion phase (3
out of 6). Since the numerical merger models do not consider
small-scale perturbations such as clumps, these galaxies might
correspond to Toomre-unstable disks driven by internal instabilities,
rather than major mergers. The post-fusion fraction of merger and
merger rate might therefore be affected by a 5.5\% and 3.3\% effect
downward, respectively (see Tab. \ref{rates}). The impact on the
pre-fusion and relaxation phases is marginal (one object) compared to
other uncertainties, so we did not consider it further.

Regarding the relaxation phase, we added six objects corresponding to
the post-fusion phase but which were not safely identified as mergers.
This could be due to insufficiently long evolution times, so that the
objects could not be safely identified, or to a too limited
exploration of the parameter space. Considering these six galaxies as
possible interlopers for the relaxation phase, this would impact the
corresponding fraction of merger and merger rate by a 9.1\% and 5.1\%
effect downward, respectively (see Tab. \ref{rates}).

\begin{figure*}
\centering \includegraphics[scale=0.65]{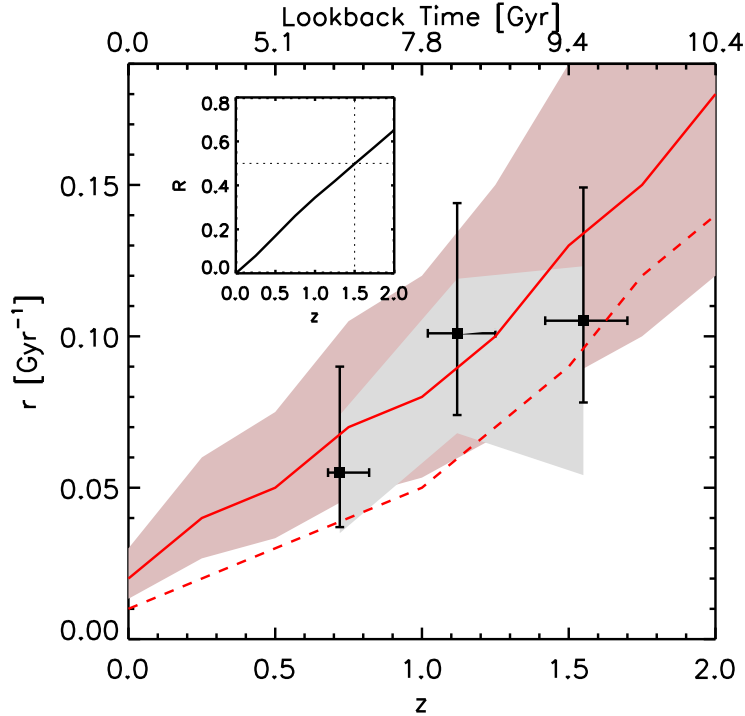}
\caption{Galaxy merger rate out to z$\sim$2. The black points with
  horizontal and vertical error bars show the observational estimates
  of the merger rate r from the IMAGES-CDFS sample. The grey-shaded
  region shows the impact of possible systematic effects of the
  observational estimation of the merger rate (see Tab. \ref{rates}).
  The red curve shows the prediction of the semi-empirical model
  \citep{hopkins09}, for the range of baryonic mass and mass fraction
  observed. The shaded red region shows the random uncertainty
  associated to the theoretical prediction (see Sect. 5.2). The
  dashed-red line shows the impact of systematic uncertainties in
  stellar mass stellar population models on model predictions (see
  Sect. 5.2). The upper-left inset shows the cumulative merger rate R
  as a function of lookback time (i.e., from the full-red curve of the
  main panel).\label{figrate}}
\end{figure*}

Systematic uncertainties are shown as a grey region in Fig.
\ref{figrate}.

\section{Comparison with semi-empirical models}

Instead of trying to predict galaxy properties all the way from
initial conditions given by the CMB, semi-empirical models by-pass the
unknown coupling between dark and baryonic matter using empirical
relations, such as gas fraction or stellar mass functions as a
function of redshifts. Such an approach allows us to overcome several
issues encountered with semi-analytic models, which can lead to
significantly underestimate the merger rate \citep{hopkins10b}. The
semi-empirical model used here \citep{hopkins09} was shown to
reproduce a number a local galaxy properties, such as their
bulge-to-disk flux ratio distribution, Tully-Fisher relation, or
luminosity function of luminous infrared galaxies \citep{hopkins10c}.

\subsection{Model parameters}
We used the \cite{hopkins09} semi-empirical model\footnote{available
  at https://www.cfa.harvard.edu/$\sim$phopkins/Site/\\mergercalc.html
  as an IDL code.} to predict the merger rate in a given range of
baryonic mass [$M_b^{min}(z)$,$M_b^{max}(z)$], gas fraction
[$f_{gas}^{min}(z)$, $f_{gas}^{max}(z)$], and baryonic mass ratio
between the two progenitors [$\mu _b^{min}(z)$,$\mu _b^{max}(z)$] as a
function of redshift $z$. Indeed, \cite{hopkins09,hopkins09c}
emphasized that the most appropriate mass would be the tightly bound
material that survives stripping to strongly perturb the primary,
which is found to be the baryonic plus tightly bound dark matter mass
within a few disk scale lengths. The best observable proxy for this
quantity is the baryonic mass \citep{hopkins09}, which was therefore
used in this study. In most studies, the galaxy merger rate is
estimated in a fixed range of stellar or baryonic mass, constant with
redshift, i.e.,
[$M_b^{min}(z)$,$M_b^{max}(z)$]=[$M_b^{min}$,$M_b^{max}$]. Here, we
need to estimate the proper range of baryonic mass
[$M_b^{min}(z)$,$M_b^{max}(z)$] at all $z$ in order to sample the
progenitors and descendants of the IMAGES-CDFS galaxies.

Given [$M_b^{min}(z)$,$M_b^{max}(z)$], we determined the corresponding
range of halo mass [$M_h^{min}(z)$,$M_h^{max}(z)$] using the
relationship between baryonic and halo masses $M_b-M_h$ as provided by
the \cite{hopkins09} halo occupation model (HOD) using the mass
function from \citet{perez09}. Using alternative mass functions change
the merger rate by less than 1\%. We then estimate the average mass of
the progenitor halos at $z+dz$ following \cite{neistein06,neistein08},
i.e., [$M_h^{min}(z+dz)$,$M_h^{max}(z+dz)$], which is then converted
back to [$M_b^{min}(z+dz)$,$M_b^{max}(z+dz)$] using the $M_b$-$M_h$
relationship. The HOD model is used to estimate the galaxy merger rate
at $z+dz$ in the updated range of baryonic mass. This ensures that the
derived galaxy merger rate is integrated over the population of the
progenitors of all galaxies within [$M_b^{min}(0)$,$M_b^{max}(0)$] at
$z=0$ only, and not over all the mass spectrum, which would
overestimate the merger rate compared to estimates from observations.
In practice, we set [$M_b^{min}(0)/M_\odot$,
$M_b^{max}(0)/M_\odot$]$\simeq$[$10^{10.4}$,$10^{11.1}$] in order to
sample the range
[$M_b^{min}(z=0.6)/M_\odot$,$M_b^{max}(z=0.6)/M_\odot$]=[$10^{10.2}$,$10^{11.0}$],
which corresponds to the baryonic mass range of the IMAGES-CDFS sample
(see Sect. 2). We tested that following the mean progenitor mass
instead of the average progenitor mass between the upper and lower
limit of the mass range does not change the resulting merger rate
significantly.

To test the consistency of the adopted HOD model with the IMAGES
sample, we checked that, according to the HOD model, the range of
baryonic mass sampled at z=0.6 corresponds to a range in
$\log{(M_{stellar}/M_\odot)}$=[$10^{10.0}$,$10^{11.0}$], which also
matches the range of observed stellar mass in the IMAGES sample (see
Sect. 2.1). At z=0, the HOD model predicts that the z=0.6 progenitors
as defined by their range of baryonic mass as above, will end up in a
stellar mass range of
$\log{(M_{stellar}/M_\odot)}$=[$10^{10.3}$,$10^{11.1}$], corresponding
to the baryonic mass range of
$\log{(M_{b}/M_\odot)}$=[$10^{10.4}$,$10^{11.1}$] as quoted above. The
evolution in stellar mass as predicted by the HOD model is therefore
$\sim$0.1-0.3 dex.

The median gas fraction in the progenitors of the IMAGES-CDFS sample
were estimated using their observed SFRs \citep{hammer09b}. We used
this to parameterize the expected gas fraction in the IMAGES-CDFS
progenitors and descendants, using
$f_{gas}$=\{15\%,32\%,50\%,75\%,90\%\} at
$z$=\{0,0.6,0.72,1.12,1.55\}. Estimates at $z$=0 and 0.6 were derived
using local HI measurements and inversion of the Schmidt-Kennicutt
relation, respectively \citep{puech10,rodrigues11}. Estimates at
$z$=0.72, 1.12, and 1.55 were derived from the estimated median gas
fractions in the pre-fusion, post-fusion, and relaxation phase,
respectively \citep{hammer09b}. The HOD model predicts that the
progenitors of z$\sim$0.6 on-going mergers in the IMAGES-CDFS sample
are galaxies with stellar mass in the range
$\log{(M_{stellar}/M_\odot)}\sim$ [$10^{9.2}$,$10^{10.4}$]. Gas
fractions of the order of $\sim$90\% in such a range of mass at
z$\sim$1.5 is consistent with observational estimates of the gas
fraction in more z$\sim$1.5 massive galaxies (see discussion in Sect.
6.2). We added a $\pm$1-$\sigma$ scatter of 0.15 dex, which appears to
represent well the observed scatter at all $z$
\citep{hopkins09,hopkins09b,rodrigues11}. At given $z$, the range of
gas fraction [$f_{gas}^{min}(z)$,$f_{gas}^{max}(z)$] over which the
merger rate was integrated is interpolated from the $f_{gas}(z)$
relationship as defined above, with an added $\pm$2-$\sigma$ scatter.
We explored the impact of a larger $\pm$3-$\sigma$ scatter with no
qualitative change in the results. Finally, we set [$\mu
_b^{min}(z)$,$\mu _b^{max}(z)$]=[$\mu _b^{min}$,$\mu
_b^{max}$]=[0.25,1.0], as found in the IMAGES-CDFS sample
\citep{hammer09b}.

\subsection{Comparing observations and theory}
Figure \ref{figrate} compares the derived estimates of the merger rate
with state-of-the-art semi-empirical $\Lambda$-CDM models
\citep{hopkins09}. This theoretical model is found to be in remarkable
agreement with the observed estimates without any particular
fine-tuning. At the redshifts and masses probed by the IMAGES-CDFS
sample, semi-empirical (HOD-based) models can predict the merger rate
within a factor $\sim$1.5 \citep{hopkins10b}. HOD models are in
relatively good agreement one with each other because they rely on
observational constraints (e.g., the galaxy stellar mass function)
that are well constrained at z$\leq$1 and for stellar masses around
$M^*$ \citep{hopkins10b}. As stated, above, changing the galaxy
stellar mass function changes the resulting merger rate by less than
$\sim$1\%. The dominant source of random uncertainty is therefore due
to variations from one (HOD) model to another, which is illustrated in
Fig. \ref{figrate} as a red region. This is the same order of
magnitude that random uncertainties associated to the observed merger
rate (see Tab. \ref{rates}). This implies that random uncertainties
limit the comparison between observations and theory at a level of a
factor $\sim$2.

In this comparison, we adopted a diet Salpeter IMF for all observables
at any epoch. Several studies suggest that the IMF might become
top-heavy under extreme starburst conditions \citep{weidner10}. This
could result in an evolution of the IMF with redshift, which would
impact the estimation of stellar mass by up to $\sim$ -0.18 dex
(rising almost linearly with redshift from z=0 to z$\sim$1.5) compared
to estimates at z=0 \citep{vandokkum07}. This could result in a
systematic shift of the merger rate with redshift. Several systematic
effects associated with the stellar population models can also affect
the determination of the stellar mass. The most significant effect is
the influence of the TP-AGB phase, which omission in stellar
population models as those used in the present study can lead to
overestimate the stellar mass by 0.14 dex \citep{pozzetti07}. In
addition, using simplified prescriptions for deriving the stellar-mass
instead of full SED modelling \citep{bell03}, when used at high
redshifts, can also potentially lead to overestimate stellar masses by
0.2 dex \citep{puech08,gallazzi09}.

In order to bracket the upper limit of the predicted merger rate at
all redshifts, we combined quadratically the systemic uncertainty
associated with a possible evolution of the IMF, and the effects of
stellar population models, since they are independent. This results in
a total systematic uncertainty of -0.27 dex on the baryonic mass. In
Fig. \ref{figrate}, we plotted theoretical predictions with
[$M_b^{min}(z)$,$M_b^{max}(z)$]-0.27, using a dashed-red line. In
doing so, [$f_{gas}^{min}(z)$,$f_{gas}^{max}(z)$] was updated
accordingly (upward). All these systematic uncertainties associated to
the estimation of the stellar mass tend to decrease the galaxy merger
rate at all redshifts. However, the theoretical predictions remain
consistent with the observational estimates within the corresponding
range of observational systematic uncertainties. Quantitatively, this
could decrease the integrated merger rate by $\sim$20\%.

We conclude that the major merger rate is determined from both theory
and observations within a factor 2-3, accounting for random and
systematic uncertainties. Such an agreement between the merger rate
determined using morpho-kinematic data and theoretical predictions
without any particular fine-tuning, is remarkable. This is an
improvement of a factor $\sim$3 compared to previous comparisons
between observations and theory \citep{hopkins09}.

\section{Discussion}

\subsection{One third of z$\sim$0.6 galaxies are on-going major mergers}
Using morpho-kinematic data of z$\sim$0.6 galaxies from the
IMAGES-CDFS survey, we derived the evolution of the major merger rate
out to z$\sim$1.5, i.e., nine billion years ago. Using comparisons of
the spatially-resolved morpho-kinematic properties between
observations and a grid of hydrodynamical simulations of galaxy
mergers, we constrain their time evolution, which allowed us to
accurately date back when they were still in pair. In this comparison,
the estimate of the merger rate at z$\sim$1.5 is much more uncertain
since it relies on the relaxation phase during which it is more
difficult and/or degenerate to interpret morpho-kinematic features as
merger relics. However, it makes sense to interpret such galaxies as
post-mergers in the relaxation phase as detailed above, and Fig.
\ref{figpot} \& \ref{figtime} indeed suggest a plausible causal link
between distant merging galaxies and such dynamically hot rotating
disks. We found a remarkable agreement between these new observational
estimates and predictions from state-of-the-art semi-empirical
$\Lambda$-CDM models within a factor 2-3, accounting for both random
and systematic uncertainties.

As a test of consistency, the $\Lambda$-CDM model predicts a large
fraction of peculiar galaxies at z=[0.4-0.75], which is found to be
$<r>$ (i.e., the average merger rate at z=0.65, $\sim$0.06 Gyr$^{-1}$,
see Fig. \ref{figrate}) $\times~\tau$ (i.e., the total duration of the
average major merger at z=0.65, $\sim$5.5 Gyr, see Fig. \ref{figtime})
$\sim$ 33\%. This corresponds exactly to the fraction of observed
galaxies in which the peculiar morphologies and/or kinematics can be
reproduced by a major merger model (see Sect. 3.2). The $\Lambda$-CDM
model therefore predicts that $\sim$33\% of the progenitors of local
disks were destroyed over the past 9 Gyr during major (i.e., baryonic
mass ratio $\geq$1:4) merger events, as observed.

The occurrence of 33\% of gas-rich major mergers in z$\sim$0.6
galaxies is also consistent with an increase of the scatter of the
Tully-Fisher relation at high redshifts
\citep{flores06,kassin07,puech08,covington10,puech10} \footnote{It has
  been recently claimed that the distant stellar mass TF relation
  might have a scatter that is actually similar to the local relation
  (Miller et al. 2011). The reduced scatter has been attributed to the
  long integration times that would better sample the plateau of the
  rotation curve combined to an optimal velocity extraction method,
  providing a more accurate circular velocity estimation. We note that
  exposure times in the IMAGES survey are comparable or even larger
  that those used in this study. Moreover, IFU data provide additional
  element of resolutions in the direction perpendicular to the galaxy
  major axis, which provide more information to accurately deconvolve
  the line-of-sight velocity measurement and estimate the rotation
  velocity. This is essential when dealing with galaxies with complex
  VFs, since the kinematic and morphological axes can be significantly
  misaligned, or when deconvolving the measurement from seeing
  effects. It is beyond the scope of this paper to provide a
  definitive assessment on the evolution of the scatter of the TF
  relation with redshift.}, and the observed random walk in the
specific angular momentum of z$\sim$0.6 galaxies compared to local
galaxies \citep{puech07}.

\subsection{Disk rebuilding of local bulge-dominated spirals}

The agreement found between observations and theoretical predictions
(see Fig. \ref{figrate}) evidences that 50\% (65\%) of local galaxies
with stellar mass in the range 10$^{10.3}$-10$^{11.1}$M$_\odot$ (see
Sect. 5.1) have undergone a major merger (defined here as baryonic
mass ratio $\geq$0.25) since z=1.5 (z=2.0). The \cite{hopkins09} model
takes into account the specific physics involved in gas-rich major
mergers, which was shown to reproduce the observed local relation
between bulge-to-disk ratio and stellar mass
\citep{hopkins09,hopkins09b}. This suggests that the time at which
local galaxies have undergone their last major merger could be an
important factor in determining how they acquired their internal
structure (B/T ratio and angular momentum).

Depending on the gas fraction at fusion time, major mergers are
well-known to result in ellipticals (through dry mergers), but also in
spirals (through gas-dominated mergers). Indeed, hydrodynamical
simulations of isolated merging disk galaxies have shown that with a
fraction of gas at the fusion time larger than 50\%, a significant
disk can reform in the remnant
\citep{springel05,robertson06,hopkins09c}. Note that the required
fraction could be smaller with significant cosmological gas accretion
\citep{brooks09}, but we will adopt 50\% as a conservative limit.

The median gas fraction at z=0.6 is found to be $\sim$30\%
\citep{puech10,rodrigues11}. It is therefore likely that major mergers
occurring between z$\sim$0.6 and z=0 (which involved 20\% of local
galaxies according to Fig. \ref{figrate}) were relatively dry,
probably resulting in galaxies with significant bulges.

The major mergers sampled by the IMAGES-CDFS sample are those occurring
between z$\sim$1.5 and z$\sim$0.6 (see Tab. \ref{rates}). Fig.
\ref{figrate} shows that 50\% of local galaxies were involved in a
major merger since z=1.5, of which 30\% occurred between z=0.6 and
z=1.5. \cite{hammer09b} have extrapolated the gas fraction in the
progenitors of these mergers, and found gas fractions well in excess
of the required value of 50\% to reform a disk ($\sim$90\%, see Sect.
5.1). Several measurements of the gas fraction were attempted at
z$\sim$1.5 \citep{daddi10,weinzirl11}. Most measurements involve
relatively massive galaxies
($\sim$M$_{stellar}=10^{10.7}-10^{11.1}$M$_\odot$) and are limited to
their half light radius. However, the HOD model predicts that the
progenitors of the on-going z$\sim$ 0.6 major mergers at z=1.5 should
be in a lower range of stellar mass
$\sim$10$^{9.2}$-10$^{10.4}$M$_\odot$. \cite{rodrigues11} homogenized
gas estimates from the literature to parameterize the evolution with
redshift and stellar mass of the gas fraction within the full gas
radius. Extrapolating their gas fraction-stellar mass relation at
z$\sim$1.5 in the range of mass 10$^{9.2}$-10$^{10.4}$M$_\odot$, one
finds that gas fractions at a level of $\sim$ 80-100\% are probably
close to average conditions (see Fig. 7 of \citealt{rodrigues11}). The
progenitors of all the z$\sim$0.6 on-going major mergers at
z=[0.75-1.5] should therefore be rich enough in gas so that a
significant disk reform after fusion, as also evidenced by
observations of young dust-enshrouded disks at z$\sim$0.6
\citep{hammer09,puech09}.

We emphasize that we have considered here only the ``average'' results
of major mergers since z$\sim$2 and such statements have to be taken at
the lowest order only, since one expects significant scatter in the
B/T of a major merger remnant as a function of orbital parameters, gas
fraction, and feedback \citep{hopkins08,hopkins09,governato09}. We
also emphasize that we considered here only a specific range of mass
and that the impact of more modest mergers, i.e., with baryonic mass
ratio smaller than 0.25 was not considered.

\subsection{Is there a disk survival issue?}

Results of semi-analytic models suggest that the likeliest channel for
the formation of disk-dominated galaxies is a relatively quiet merger
history (e.g., see discussion and references in \citealt{weinzirl09}).
Of particular interest are local ``bulgeless galaxies '' (which we
define here as galaxies with B/T$\leq$0.2, as generally considered),
which are often thought to have undergone their last major merger
above z$\sim$2. The relatively large fraction of such galaxies is
often considered to be in tension with the fraction of major mergers
as inferred from predictions of $\Lambda$-CDM models (e.g.,
\citealt{stewart08}). It was argued in Sect. 6.2 that most descendants
of IMAGES-CDFS galaxies should lie within early-type spirals. We now
examine whether the fraction of mergers found in Fig. \ref{figrate} is
in tension with the observed fraction of local bulgeless galaxies.

\cite{weinzirl09} used Bulge+Bar+Disk decompositions on H-band images
and found that two-third of local spirals with
M$_{stellar}$=10$^{10.1}$-10$^{11.6}$M$_\odot$ are bulgeless. Their
determination maximizes the fraction of bulgeless galaxies since they
removed the contribution of the bar to the central light and followed
the common assumption that bars are pure products of secular evolution
and not merging (see discussion in \citealt{athanassoula08}). Since
the fraction of spirals amongst local galaxies is $\sim$70\% (see App.
B), this would translate into a maximal fraction of bulgeless galaxies
in the local Universe $\sim$46\%. As \cite{weinzirl09} argued, their
sample appears to be representative of the B-band luminosity function
at M$_{stellar}\geq$10$^{10.1}$M$_\odot$(see their Fig. 2). However,
their Fig. 3 evidences that their sample can be representative of
stellar mass function only above
M$_{stellar}\geq$10$^{10.6}$M$_\odot$. In this range of mass, the
maximal fraction of local bulgeless is found to be 61\%, or 43\% of
local galaxies. Assuming that such galaxies cannot be formed by
mergers, this would therefore limit the fraction of major merger in
local galaxies since z=2 to 57\%. We re-derived Fig. \ref{figrate} for
the range of mass M$_{stellar}$=10$^{10.6}$-10$^{11.6}$M$_\odot$.
Indeed, the HOD model used in the present study predicts that on-going
major merger in the z$\sim$0.6 IMAGES-CDFS sample should end up in a
stellar mass range of $10^{10.3}$-$10^{11.1}$M$_\odot$, which is
shifted to smaller mass relative to the range
$10^{10.6}$-$10^{11.6}$M$_\odot$ in which the \cite{weinzirl09} sample
is representative. In this higher range of mass, we found that the
fraction of local galaxies that have undergone a major merger since
z=2 decreases from 65\% to 53\%. Within uncertainties, this matches
quite well the maximal fraction of bulgeless galaxies found by
\cite{weinzirl09} in the same range of mass (57\%).


We conclude that there is no tension between the fraction of mergers
inferred from morpho-kinematic data and the fraction of local
bulgeless galaxies. In other words, there is no need for disks to
“survive” major mergers in the $\Lambda$-CDM paradigm, at least at
z$\leq$2.

\subsection{Disk rebuilding in the cosmological context}
\emph{The results of this paper point toward a re-building of half of
  local disks during the past 9.4 Gyr after their last gas-rich major
  merger \citep{hopkins09,hopkins09b,hammer05,stewart09}.} This does
not imply that the evolution of the star formation and stellar mass
densities are primarily driven by such processes over the same period
\citep{robaina09,hopkins10c}. These results are consistent with recent
developments of the $\Lambda$-CDM model at earlier lookback times
(i.e., z$\geq$1.5), according to which galaxy formation would be
mainly driven by streams of cold gas, rather than by major mergers
\citep{forster09,agertz09,weinzirl11}. Indeed, cosmological
simulations predict that cold flows are strongly suppressed below
z$\sim$1-1.5 \citep{keres09}. Such an epoch might correspond to a
transition between two different mechanisms driving galaxy formation,
i.e., cold gas streams and gas-rich major mergers, as suggested by the
strong evolution of the fraction and properties of clumpy galaxies as
a function of redshift \citep{puech10b,contini12b}. In such
simulations, halos with a relatively quiet merger history are usually
selected as the best probes for the formation of local thin disks
(e.g., \citealt{brook11}). The present results suggest that gas-rich
major mergers could be be an equally plausible way of forming large
thin disks. Further work will be needed to quantify what is the main
reservoir of gas from which such thin disks can reform. Potential
sources of gas include gas expelled during the merger and which falls
back onto the remnant \citep{barnes02}, gas coming from external gas
accretion \citep{brooks09}, and/or gas recycled through stellar
feedback \citep{martig10}. This emphasizes the need to further explore
how such disks can (re-)form in halos with a more recently active
merger history \citep{governato09}. Interestingly, recent cosmological
simulations conducted with state-of-the-art hydrodynamical codes
reveal cold gas disks that have larger extent and specific angular
momentum compared to disks formed in former SPH cosmologicial
simulations \citep{torrey11,keres11}. These simulations show that
massive halos (i.e., $M_{halo}\geq 10^{11}M_\odot$) have significantly
larger SFRs at z=0-2, compared to previous simulations
\citep{keres11}.



\section{Conclusion}
Using morpho-kinematic data of z=[0.4-0.75] galaxies from the
IMAGES-CDFS survey, we derived the evolution of the major merger rate
out to z$\sim$1.5, i.e., nine billion years ago. Using comparisons of
the spatially-resolved morpho-kinematic properties between
observations and a grid of hydrodynamical simulations of galaxy
mergers, we constrain their time evolution, which allowed us to
accurately date back when they were still in pair. We then compared
constrains from observations with predictions from semi-empirical
models. Our main conclusions are as follows.

\begin{enumerate}
\item Morpho-kinematic observations are found to be sensitive to all
  the phase of the merger process, from the pre-fusion phase in which
  galaxy nuclei are still separable by visual inspection, to the
  post-fusion and relaxation phases, in which the two progenitors
  merge and progressively reach equilibrium. Each of the three phases
  are found to last $\sim$1.8 Gyr.

\item The major merger rate (defined as baryonic mass ratio
  $\geq$0.25) inferred from morpho-kinematic observations is found to
  be evolving from 5.5\%.Gyr$^{-1}$ to 11.1\%.Gyr$^{-1}$ between
  z=0.72 to z=1.55 in the range of stellar mass
  10$^{10}$-10$^{11}$M$_\odot$ (diet Salpeter IMF), or
  10$^{10.2}$-10$^{11}$M$_\odot$ in baryonic mass.

\item A remarkable agreement is found between observational estimates
  and predictions from the \cite{hopkins09} semi-empirical model,
  which accounts for the specificity of gas-rich mergers at high
  redshifts. The agreement is found to be at a level of a factor 2-3
  without any particular fine-tuning of the model, accounting for both
  random and systematic uncertainties on the observational and
  theoretical sides, which significantly improves the accuracy of
  previous comparisons.

\item The $\Lambda$-CDM model predicts, in agreement with
  observations, that 33\% of z$\sim$0.6 intermediate-mass galaxies are
  on-going major mergers. Observational signatures of these mergers
  encompass morpho-kinematic disturbances in a large fraction of these
  galaxies at this redshift, and an increase of the scatter of scaling
  relations between mass, velocity and radius (or angular momentum).

\item The $\Lambda$-CDM model predicts, in agreement with
  observations, that 20\% / 50\% / 65\% of local galaxies in the range
  of stellar mass 10$^{10.3}$-10$^{11.1}$M$_\odot$ were involved in a
  major merger between z=0.6 / 1.5 / z=2.0 and z=0.

\item Most of the major mergers occurring between z=0.6 and z=0 were
  probably relatively dry, resulting in galaxies with a prominent
  bulge, while most of the major mergers occurring between z=1.5 and
  z=0.6 probably involved large gas fractions, which should lead to
  the rebuilding of a significant disk.

\item Theoretical models often suggest that ``bulgeless galaxies''
  (B/T$\leq$0.2) result from relatively quiet merger histories. We
  found no tension between the fraction of local bulgeless galaxies
  and the fraction of local galaxies that have undergone a major
  merger since z=2, i.e., there seems to be no ``disk survival issue''
  but rather a ``disk rebuilding'' era, at least at z$\leq$1.5-2, when
  the influence of cosmological gas accretion predicted by theory and
  simulations would start declining in relatively massive haloes.
\end{enumerate}

\acknowledgments This work was partly supported by the ``Laboratoire
International Associ\'e'' Origins. We thank an anonymous referee for
her/his careful reading of the paper and thorough comments.

\appendix

\section{Examples of morpho-kinematics data and associated major merger models in the IMAGES-CDFS sample }
We present here the i-band HST/ACS images and kinematic maps for a
sub-sample of the IMAGES-CDFS, see Fig. \ref{vf}. Galaxies are the
same that those used to illustrate the SFR vs. time sequence of Fig.
\ref{figtime}, in the same (i.e., temporal) order, from top to bottom.
Galaxies with no major merger model are RDs.

\begin{figure*}[p]
\centering \includegraphics[width=15cm]{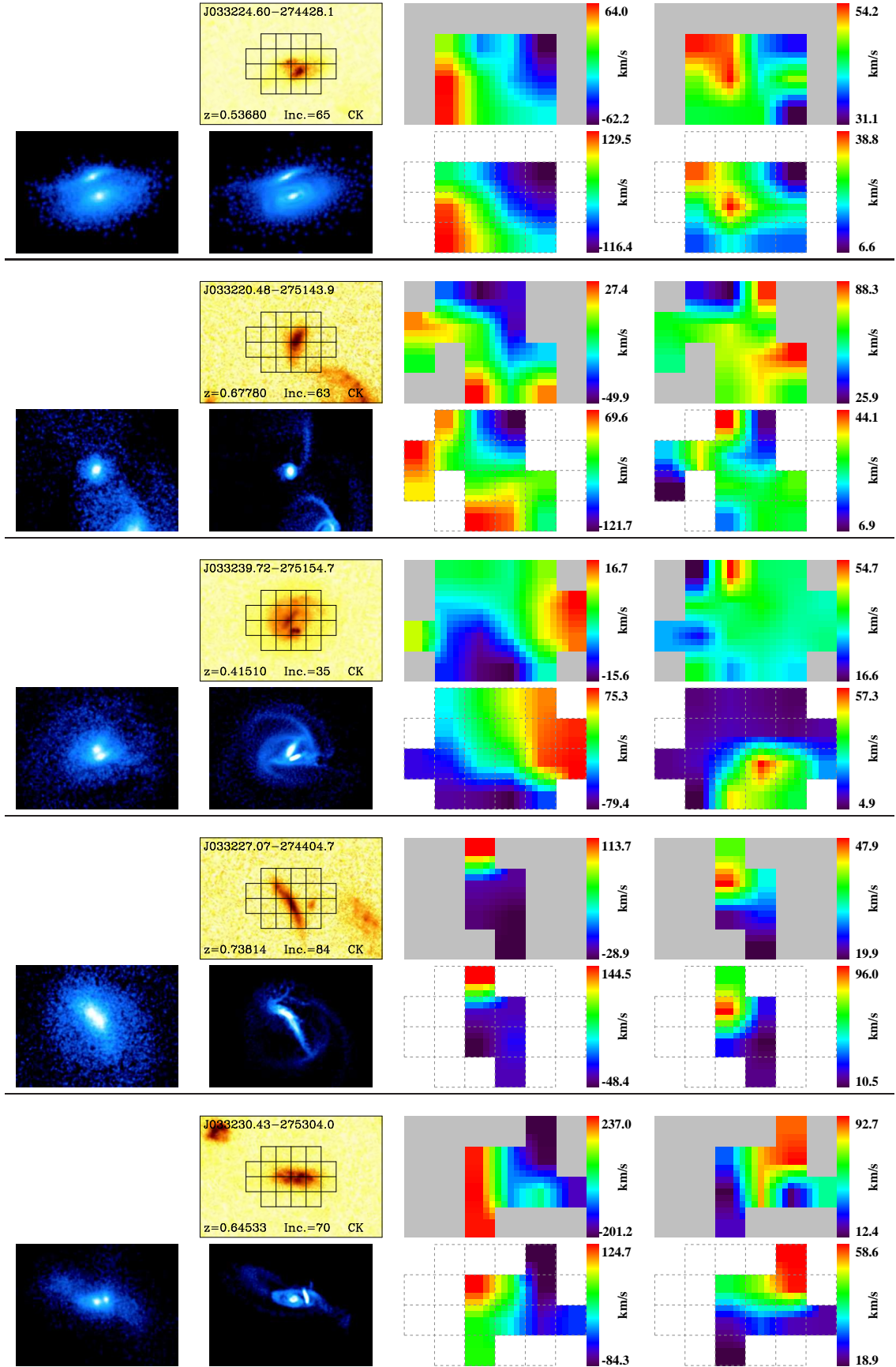}
  \caption{Morpho-kinematic observations for an IMAGES-CDFS subsample
    and results of their modelling. See text for details. For each
    galaxy, from left to right: HST/ACS i-band image with the GIRAFFE
    IFU superimposed, GIRAFFE velocity field, and GIRAFFE velocity
    dispersion map (top panel), modelled stellar density, modelled gas
    density, simulated velocity field, and simulated velocity
    dispersion map (bottom panel). See \cite{hammer09b} for details.}
\label{vf}
\end{figure*}

\setcounter{figure}{6}
\begin{figure*}
\centering \includegraphics[width=15cm]{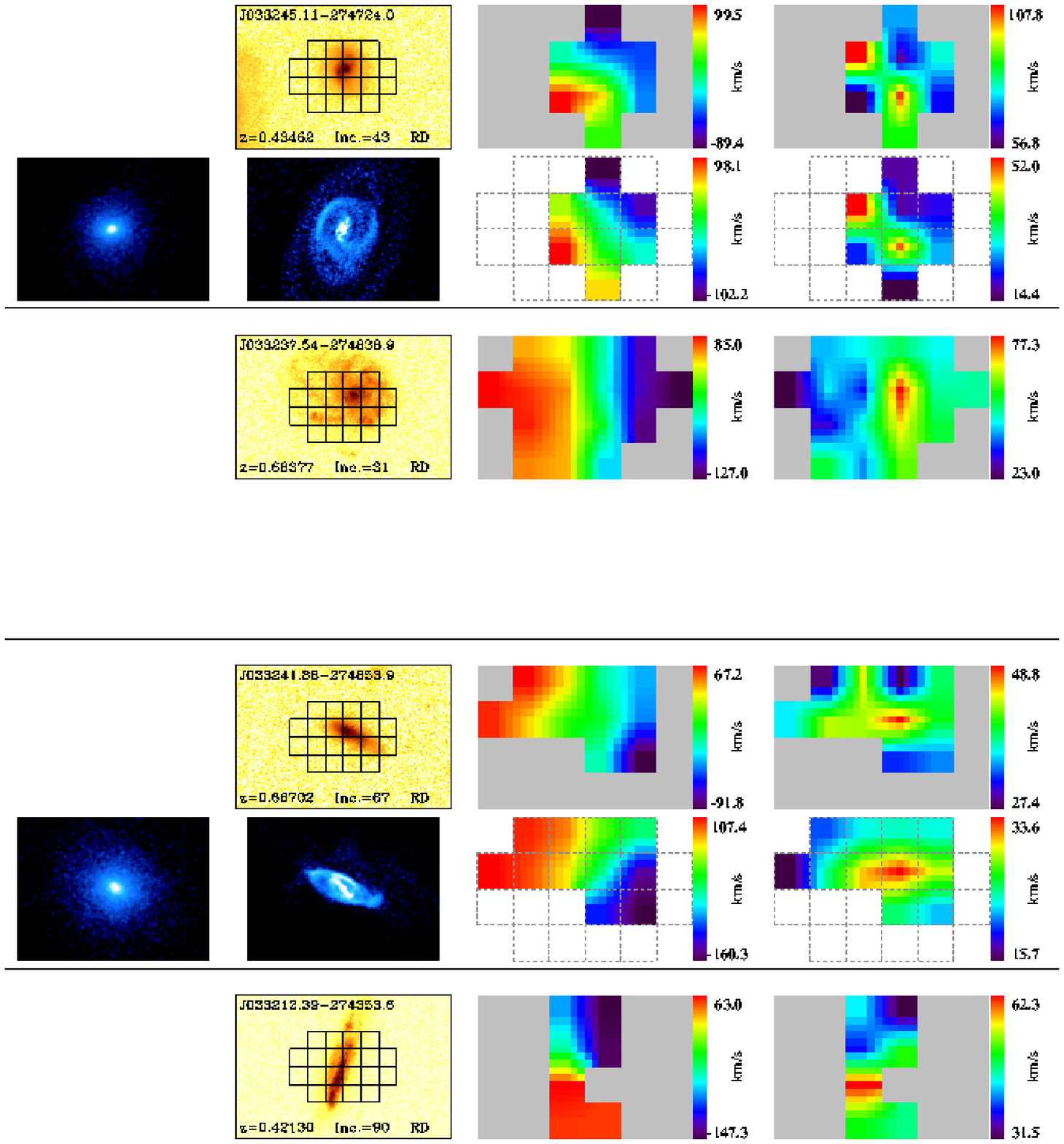}
  \caption{Continued.}
\end{figure*}

\section{Evolution of galaxy morphology between z$\sim$0.6 and z=0}

We discuss here the evolution of the fraction of morphological types
between z$\sim$0.6 and z=0 in the range of stellar mass which
encompasses the descendant of the IMAGES-CDFS galaxies.

We estimated the fraction of each morphological type in local galaxies
as follows. We first considered Morphologically Determined Luminosity
Functions (MDLF) of local galaxies from \cite{nakamura03}. They indeed
constructed LFs for E/S0, S0/a-Sb, Sbc-Sd, and Im galaxies using
$\sim$1500 galaxies from the SDSS using the r band luminosity. We
first converted their observed r band luminosity into stellar mass
using prescriptions from \cite{bell03}, which gives a mean stellar
mass/r-band luminosity ratio in the local volume of 3.05 within
$\pm$30\% of systematic uncertainty (diet Salpeter IMF). We then
integrated each of the MDLF (once converted into the proper IMF) from
10$^{10.3}$ to 10$^{11.1}$M$_\odot$, within which the descendants of
the on-going z$\sim$0.6 major mergers should end up, as predicted by
the HOD model (see Sect. 5.1). This results in the fractions of
morphological types as listed in Tab. \ref{types}. Deriving the
fraction of morphological type that way has the advantage of avoiding
systematics associated with the representativity of the sample, and is
limited only by the statistical scatter associated with the visual
classification of \cite{nakamura03} and the systematic uncertainty
associated to the mean stellar mass-to-light ratio in the local volume
as determined by \cite{bell03}.

As a consistency check, we also derived the fraction of morphological
types in the local Universe using the catalogue of $\sim$14000
visually classified galaxies in the SDSS from \cite{nair10}. We
selected in their catalogue galaxies in the range of mass
10$^{10.3}$-10$^{11.1}$M$_\odot$, and with redshift $z \leq$ 0.05.
According to their Fig. 1, this select a volume-limited sample in the
stellar mass range of interest for our purposes. \cite{nair10}
emphasized that such a cut in $z$ will remove some of the brightest,
most massive galaxies in the local volume, but these galaxies are all
above the range of mass of interest, so this will not impact the
morphological fractions in this range. Derived fractions are also
listed in Tab. \ref{types}. It is interesting to compare them with
those derived by integrating the MDLFs in the local volume, since both
derivations rely on completely different selection effects. We also
listed results from \cite{delgado10}, which selected a sample of the
catalogue from \cite{fukugita07}, who visually classified a sample of
more than 2000 galaxies from the SDSS. Their sub-sample was selected
to be volume-limited with M$_{stellar}\geq$10$^{10.3}$M$_\odot$ to
specifically match the descendants of the galaxies sampled by the
IMAGES survey.

\begin{table*}
  \caption{Fractions of morphological type at z=0 in the descendants
    of the on-going z$\sim$0.6 major mergers. Such remnants are
    expected in a range of stellar mass
    $\sim$10$^{10.3}$-10$^{11.1}$M$_\odot$. The error
    reflects the uncertainty on the conversion between r-band
    luminosity and stellar mass (see text). Fractions at z=0 are indicated as derived from \cite{nakamura03}, \cite{nair10}, and \cite{delgado10}. Note that the sum of the fractions derived from \cite{nakamura03} is slightly over 100\% because they were derived from the MDLFs. Fractions at z$\sim$0.6 were determined by \cite{delgado10}. \label{types}}
\begin{center}
\begin{tabular}{ccccc}
\tableline\tableline
Morph. type & Fraction & Fraction & Fraction & Fraction\\
            & at z=0 & at z=0 & at z=0& at z$\sim$0.6\\
            & from N03 & from N10 & from DS10 & from DS10\\\hline
E/S0     &  28$\pm$3\% & 31\% & 18$\pm$5\% & 17$\pm$3\%  \\\hline
All Sp & & & &\\
S0/a-Sd & 74$\pm$3\% & 68\% & 72$\pm$8\% & 31$\pm$7\%  \\
Early-type Sp & & & \\
S0/a-Sb &  54$\pm$4\% & 45\% & -- & -- \\
Late-type Sp & & & &\\
Sbc-Sd  & 20$\pm$1\% & 23\% & -- & --  \\\hline
Im / Pec    &  2$\pm$1\% & $<$1\% & 10$\pm$3\% & 52$\pm$9\% \\\hline
\tableline\tableline
\end{tabular}
\end{center}
\end{table*}

The comparison between these three different determinations of the
fraction of morphological types in the local Universe \emph{in the same
  mass range} reveals interesting features. First, the fraction of
spirals is found to be in remarkable agreement, around $\sim$70\%.
There are some variations within the fractions of early-type spirals,
probably reflecting the uncertainty associated to eyeball
classification between S0 and S0a galaxies. The second interesting
feature is the smaller fraction of E/S0 found by \cite{delgado10} in
conjunction with their higher fraction of Peculiar galaxies compared
to other catalogues. This was attributed by \cite{delgado10} to a
number a factors that include (1) their systematic use of color maps
to identify unusual color gradients, (2) the quantification of the
misalignment between bulge and disk centroids, and (3) the compactness
of galaxies, which were moved to the peculiar class if too compact to
be decomposed into bulge and disk. Galaxies harboring such features
were classified as Peculiar by \cite{delgado10}, but are probably in
the E/S0 class in other catalogues that do not take into account such
an information. This suggests that $\sim$10\% of the E/S0 classified
by \cite{nakamura03} or \cite{nair10} are actually perturbed and not
fully relaxed. Such an interpretation is consistent with the fraction
($\sim$30\%) of early-type galaxies showing negative color gradients
(``blue-cored'' early-types) in the local Universe in the SDSS
\citep{suh10}.

At face value, the fraction of late-type spirals in Tab. \ref{types},
i.e., 20-23\% might appear in conflict with the fraction of bulgeless
spirals derived by \cite{weinzirl09}, which is found to be $\sim$60\%
(see Sect. 6.3). We recall that ``bulgeless'' means here B/T$\leq$0.2
and that the correlation between B/T and morphological type has very
large scatter (see, e.g., Fig. 14 of \citealt{weinzirl09}). Selecting
galaxies with morphological types in the range Sbc-Sd in the sample of
\cite{weinzirl09} leads to a fraction of late-type galaxies $\sim$33\%
(above M$_{stellar}$=10$^{10.6}$M$_\odot$). Since spirals account for
$\sim$70\% of local galaxies (Tab. \ref{types}), this translates into
a fraction of $\sim$23\% of late-types amongst local galaxies, which
is fully consistent with fractions listed in Tab. \ref{types}.

We also listed in Tab. \ref{types} the fraction of morphological types
found by \cite{delgado10} at z$\sim$0.6. Note that the fractions of
Peculiar galaxies, i.e., those that do not fit into the Hubble
sequence, are at both redshifts in very good agreement with those
found by \cite{vandenbergh02}. \cite{delgado10} used exactly the same
classification method and strictly equivalent data to classify local
and distant galaxies, which minimizes systematics when comparing the
two epochs. When one uses the same method, it is found that the
fraction of fully relaxed E/S0 does not evolve between z$\sim$0.6 and
z=0. Compared to other catalogues, an evolution of the fraction of
E/S0 of the order of $\sim$10\% is found, but this can be attributed
to systematics associated to the different classification schemes
used. These 10\% likely correspond to non-relaxed E/S0, as discussed
above.

\end{document}